\documentclass[08pt,twocolumn,preprintnumbers,amsmath,amssymb,nofootinbib,superscriptaddress,prl]{revtex4-1}

\usepackage{amsmath,amssymb,graphicx}
\usepackage{float}
\usepackage{latexsym} 
\usepackage{amsmath,amsthm}
\usepackage{ifpdf}
\usepackage{epstopdf}
\usepackage[dvipsnames]{xcolor}
\usepackage{epsfig}
\usepackage{dcolumn}
\usepackage{bm}
\usepackage{braket}
\usepackage[symbol]{footmisc}
\usepackage{soul}
\usepackage{xcolor}
\usepackage{graphicx}

\begin{document}
\title{Amplification of electromagnetic waves by a rotating body}
\author{M.C. Braidotti}
\affiliation{School of Physics \& Astronomy, University of Glasgow, G12 8QQ Glasgow, UK}
\author{A.~Vinante}
\affiliation{Istituto di Fotonica e Nanotecnologie - CNR and Fondazione Bruno Kessler, I-38123 Povo, Trento, Italy}
\author{M. Cromb}
\affiliation{School of Physics and Astronomy, University of Southampton, SO17 1BJ, Southampton, UK.}
\author{A. Sandakumar}
\affiliation{School of Physics and Astronomy, University of Southampton, SO17 1BJ, Southampton, UK.}
\author{D. Faccio}
\affiliation{School of Physics \& Astronomy, University of Glasgow, G12 8QQ Glasgow, UK}
\author{H. Ulbricht}
\email[Correspondence email address: ]{H.Ulbricht@soton.ac.uk; Daniele.Faccio@glasgow.ac.uk;}
\affiliation{School of Physics and Astronomy, University of Southampton, SO17 1BJ, Southampton, UK.}

\begin{abstract}
In 1971, Zel’dovich predicted the amplification of electromagnetic (EM) waves scattered by a rotating metallic cylinder, gaining mechanical rotational energy from the body.{Since then, this phenomenon has been believed to be unobservable} with electromagnetic fields due to technological difficulties in meeting the condition of amplification, that is, the cylinder must rotate faster than the frequency of the incoming radiation. Here, we {show that this key piece of fundamental physics has been hiding in plain sight for the past 60 years in the physics of induction generators. } 
We measure the amplification of an electromagnetic field, generated by a toroid LC-circuit, scattered by an aluminium cylinder spinning in the toroid gap. We show that when the Zel'dovich condition is met, the resistance {induced by the cylinder} becomes negative implying amplification of the incoming EM waves. {These results reveal the connection between the concept of induction generators and the physics of this fundamental effect that was believed to be unobservable, and hence open new prospects towards testing the Zel'dovich mechanism in the quantum regime, as well as related quantum friction effects.}
\end{abstract}

\maketitle

\emph{Introduction} -- 
Electromagnetic (EM) wave amplification from rotating media was predicted by Yakov Zel'dovich in 1971, when he showed that an incoming beam co-rotating with a rotating absorber, such as a metallic cylinder, would be reflected and amplified in the scattering process under the condition that the cylinder was spinning {so fast to see the incoming frequency of the radiation to be negative}~\cite{Zeldovich1971,Zeldovich1972,Zeldovich1986}.
Zel’dovich's prediction also showed that this mechanism would be able to amplify any quantum fluctuations of the EM field in the vicinity of the rotating body, therefore leading to a spontaneous emission of radiation from the absorbing surface. The vacuum virtual photons would be converted into real photons, draining energy from the mechanical energy of the cylinder, which would slow down its rotation~\cite{Bekenstein1998,Maghrebi2012}. \\ 
In Zel'dovich's original postulation, EM waves carrying orbital angular momentum (OAM) $\ell$ focus radially onto a spinning metallic (absorbing) cylinder. The cylinder rotation $\Omega$ must be sufficiently high so as to induce a rotational Doppler effect that shifts the EM frequency,  $\omega$, to negative values:  
\begin{equation}\label{zeldy}
    \omega - \ell\Omega<0.
\end{equation}
When this condition is satisfied, the absorption coefficient changes sign and the rotating medium loses part of its rotational energy to the outgoing waves, which are amplified.
There have been many practical proposals to verify Zel'dovich's predictions ~\cite{Bekenstein1998,Faccio2017,Gooding2020,Alu}, relying e.g. on use of a very large OAM $\ell=10,000$ or optically levitated particles spinning at GHz rates \cite{Gooding2020}, restricting the size of the system to the nano-scale~\cite{Novotny2018,TongLi2020} or on synthetic approaches \cite{Alu}.
{Despite the large quantity of proposals, technological challenges in meeting the amplification condition have prevented its verification with EM waves, making everyone believe that this long-foreseen effect was inaccessible in the near future}. The fastest rotation achievable by standard motors is of the order of $10$ kHz~\cite{celeroton}, and a record of $667$~kHz is reported for a millimeter-sized magnetically levitated sphere~\cite{Schuck2018}. {The only measurement of an analogue effect} to date has been in the acoustic regime where the speed of light is replaced by the speed of sound that is orders of magnitude lower~\cite{Cromb2020}. The key step forward in this work was the implementation of a scheme where the spinning disk is located deeply in the near-field regime, i.e. all dimensions (size of the OAM beam, radius and thickness of the spinning disk) are $\ll \lambda$. {In spite of this achievement, the acoustic analogy} does not allow an extension to the quantum regime, leaving open the challenge to observe the effect in the originally proposed EM domain. 
{Furthermore, Zel'dovich himself pointed out that with OAM the near-field approach, despite being necessary to avoid superluminal tangential velocities, leads to a very weak amplification. The EM field interacts with the cylinder in a region where the field amplitude decreases like a power of $(r/\lambda)^{\ell}$, thus reducing even further the amplification~\cite{Zeldovich1971,Zeldovich1986}. }\\
{In this paper, we show that this 60 year-old long-sought effect has been concealed for all this time in the physics of induction generators. 
Induction motors are constituted of two components: an external stator, composed of circuits generating a rotating magnetic field, and a rotor, also composed of several elementary circuit loops, usually in a squirrel cage configuration. By replacing the internal circuits of the rotor with a solid metal cylinder as in Zel’dovich's original proposal, {and using a gapped toroid within a LC resonator as stator, we isolate the key physical effect and unambiguously observe Zel’dovich amplification, which manifests itself as a negative dissipation induced  by the rotor in the LC circuit.} 
In particular, this overcomes two conceptual and  technological difficulties related to high rotational speeds and low amplification:
1) having a near-field interaction allows us to use longer EM wavelengths; 2)	using spin-rotational momentum, and not OAM, allows us to maximize the geometrical interaction area, by removing the non-wave zone intrinsic to OAM waves. 
Indeed, by exploiting the mechanical equivalence between spin and OAM for the Doppler shift of EM waves \cite{Padgett97,Padgett98}, we overcome the disadvantage of the original Zel'dovich scheme, i.e. a weak or close-to-zero intensity (decaying as $r^{|\ell|}$) in the region occupied by the rotating cylinder.} 
\begin{figure*}[t]
\centering
\includegraphics[width=14cm]{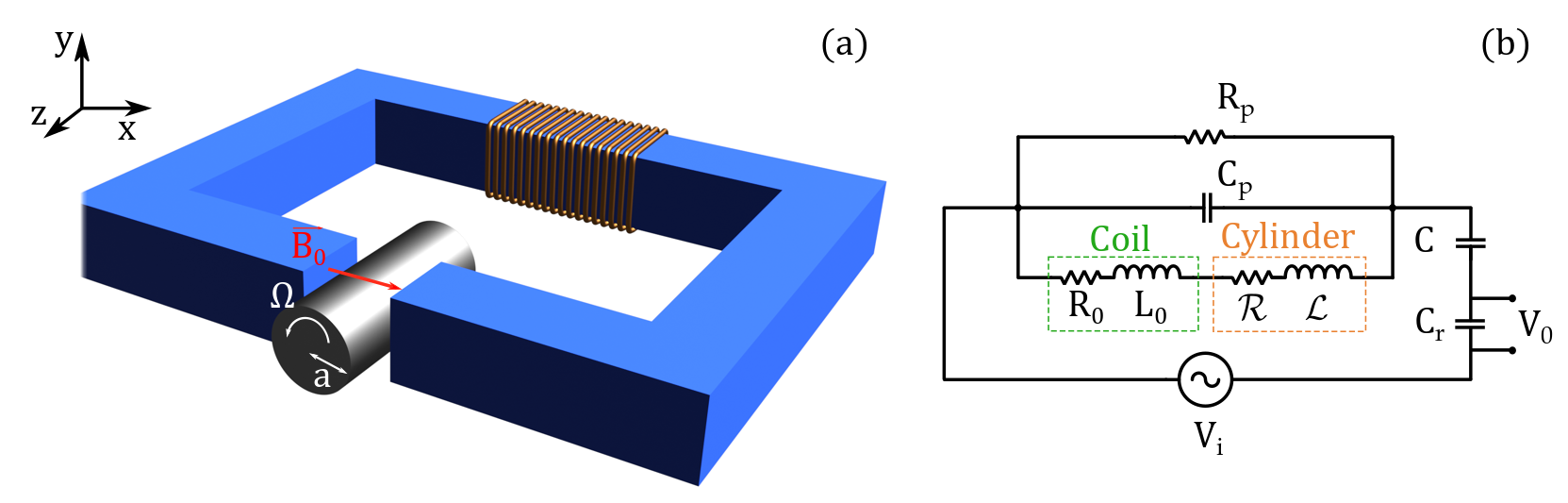}
\caption{\label{setup} (a) Experimental layout. (b) Outline of the RLC circuit used for measurements and detection. $R_0$ is the ohmic coil dissipation, while parallel dielectric losses are represented by $R_p$. 
}
\end{figure*}
\emph{The model} -- Our setup is composed of a LC toroid circuit and a spinning cylinder within the toroid gap, as shown schematially in Fig. \ref{setup}(a).\\ 
To compute the effect of the rotating cylinder on the circuit, we first calculate the currents induced by the magnetic field on the metallic cylinder, and then the field induced by the currents back into the circuit.
Following the calculation in Ref.~\cite{BraidottiPRL2020}, we consider the spinning cylinder axis oriented along $z$ and the magnetic flux density $\bm{B_0}$ produced by the coil wrapped around the toroid to be uniform and oscillating at a frequency $\omega$. 
In the lab reference frame we can write the field in the toroid gap as 
\begin{equation}
\mathbf{B_0}= \beta \mathbf{b_0}I,
\end{equation}
where $I$ is the current in the coil wrapped around the toroid. The factor $\beta$ is a geometrical factor that we measure experimentally. 
The vector $\mathbf{b_0} = (1,0,0)^T e^{i\omega t}$ indicates the linear polarization of the oscillating magnetic field. We write this as the sum of a co-rotating and a counter-rotating vector with respect to the cylinder axis, $\mathbf{b_0} =1/2 \left[(1,i,0)^T +(1,-i,0)^T \right] e^{i\omega t}$. 
We then move to the reference frame co-rotating with the cylinder, spinning at frequency $\Omega$:
\begin{equation}
\mathbf{b'_0} ={1\over2} \left[(1,i,0)^T e^{i(\omega-\Omega) t} +(1,-i,0)^T e^{i(\omega+\Omega) t}\right]. 
\label{co_cout}
\end{equation}
The response of the cylinder to the applied magnetic field is thus the superposition of the response to the co-rotating and the counter-rotating {polarization} components, which rotate at different {Doppler shifted frequencies, $\omega-\sigma\Omega$, where $\sigma$ indicates the wave spin ($\sigma=1$ co-rotating, $\sigma=-1$ counter-rotating)}. In particular,  when the condition $\Omega>\omega$ is fulfilled, the co-rotating frequency {$\omega-\Omega$ flips sign.} \\
The response of the cylinder to each component can be found by solving Maxwell's equations for the specific geometry and material. Analytical solutions can be found in the case of a spherical rotor \cite{BraidottiPRL2020} and an infinite cylinder (see  \cite{supply}). For practical purposes, the sphere solution is found to adequately describe our experiment, as shown below. For our conductive rotor, the response is determined by the eddy currents induced in the conductive material of the cylinder, involving both an inductive (in-phase) and resistive (out-of-phase) component. Induced eddy currents will couple a magnetic flux back into the coil circuit,    $\Phi\left( \omega \right) =  \alpha  \left(  \omega \pm \Omega  \right) I\left( \omega \right)$, 
where the linear susceptibility $\alpha=\alpha'- i\alpha''$ is evaluated in the rotating frame. Here, $\alpha'$ and $\alpha''$ are the in-phase and out-of-phase components of the response function. In general, $\alpha$ is proportional to the geometrical coupling $\beta^2$. 
Moreover, according to linear response theory, $\alpha \left( - \omega \right) $ = $\alpha^* \left( \omega \right) $, in particular $\alpha'' \left( -\omega \right) $ = $-\alpha'' \left( \omega \right) $. \\
%
%
%
%
From an experimental point of view it is convenient to define the resistance $\mathcal{R}$ and the inductance $\mathcal{L}$ induced by the rotating cylinder into the circuit:
\begin{eqnarray}
\label{m_yZ}
\mathcal{R} &=& Re[V/I] = \omega \left[ \alpha'' \left(  \omega - \Omega  \right)+ \alpha'' \left(  \omega + \Omega  \right) \right] \\
\mathcal{L} &=& Re [\Phi/I] = \alpha'\left(  \omega - \Omega  \right) + \alpha'\left(  \omega + \Omega  \right).
\end{eqnarray}
where $V=i\omega \Phi$ is the induced voltage.\\
The in-phase component $\alpha'$ gives the variation of inductance $\mathcal{L}$ generated by the presence of the rotating cylinder in the gap, while $\omega\alpha''$ can be seen as an induced resistance. Hence, the condition $\omega-\Omega<0$ for the co-rotating component leads to a negative resistance that in turn implies a power emission into the EM mode as opposed to the expected (for a non-rotating or slowly rotating cylinder) power absorbed from the EM mode. This corresponds to the amplification predicted by Zeldovich in free space \cite{Zeldovich1971,Zeldovich1972,Zeldovich1986}.\\
However, note that in the case of a linearly polarized oscillating field the total resistance $\mathcal{R}$ is always composed of a co-rotating and a counter-rotating component. We can have amplification only if the negative resistance induced by the co-rotating term is larger than the counter-rotating one (which is always positive). This can indeed occur due to the symmetry breaking induced by the mechanical rotation and the counter-rotating term could be completely eliminated in a setup with a circularly polarized field. \\
\begin{figure}[t]
\includegraphics[width=\columnwidth]{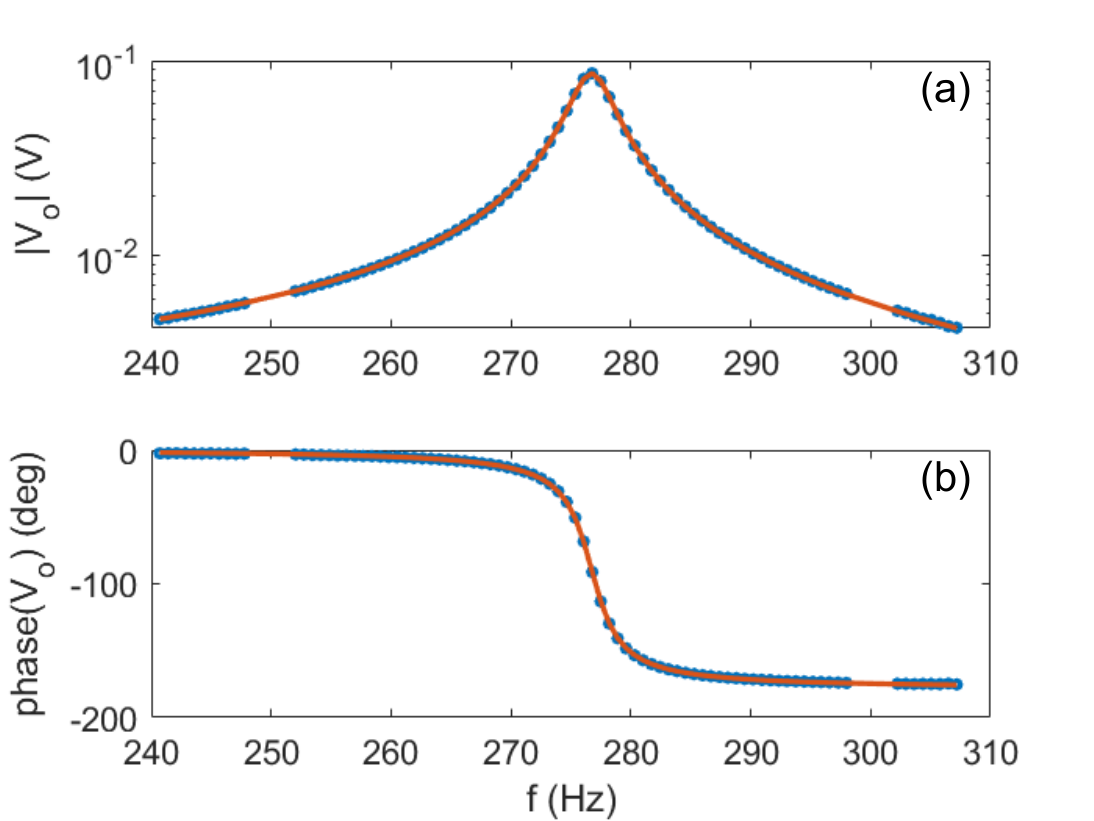}
\caption{\label{plotesempio} {Example of the amplitude response $|V_o|$ (a) and phase response $phase(V_o)$ (b) of the LC coil-capacitor setup, measured by the lock-in amplifier, as function of the generator frequency $f=\omega /\left( 2 \pi \right)$. In this example the rotor mechanical frequency is fixed at $F=\Omega/(2 \pi) = 330$~Hz and the LC resonance frequency is $f_0=277$~Hz, obtained with capacitors $C=1.0$ nF, $C_r=22$ nF. From the fit, shown as the red lines with modulus and phase of Eq. (\ref{Vo}), we can extract the parameters $L$ and $R$.}
}
\end{figure}
\emph{Experimental setup} --
In our experimental setup the rotor is an aluminum cylinder with radius $a=2$ cm, mounted on a brush-less motor \cite{maxon} which can be spun up to $500$ Hz about its symmetry axis (see Fig.~\ref{setup}(a)). The magnetic field is generated by a coil wound around a gapped toroidal ferrite core with square section $4 \times 4$ cm. The rotor is inserted in the $4.4$ cm gap, slightly larger than rotor diameter. The coil is made of $2 \times 10^4$ turns of $0.2$ mm diameter copper wire. The ohmic coil resistance at room temperature is $R_0=2.03$ k$\Omega$, while the measured coil inductance is $L_0 \approx 263$ H. We approximate the magnetic field as quasi-uniform over the gap. The current-to-field factor $\beta=(0.40 \pm 0.03)$ T/A has been directly calibrated with a Gaussmeter (Hirst GM05) placed in the center of the gap. \\
\emph{Measurement methods and data analysis} --
In order to perform accurate measurements of the resistance $\mathcal{R}$ and inductance $\mathcal{L}$ induced by the rotor, the coil is placed in series to a capacitor, thus forming a $RLC$ circuit, whose scheme is shown in Fig.~\ref{setup}(b). We measure the complex transfer function from the input voltage $V_i$, applied to the RLC circuit by a function generator, to the output voltage $V_o$, measured across a large readout capacitor $C_r$. The measurement is performed by a lock-in amplifier \cite{zurich} synchronous with the input signal. The capacitor $C_r$ has an impedance much lower than the lock-in input impedance {$R_{\mathrm{in}}=1$ M$\Omega$,} in order to suppress the parasitic dissipation induced by ${R_{\mathrm{in}}}$.\\
A typical measurement of amplitude and phase of the output voltage {$V_o$, measured by the lock-in amplifier,} as functions of the generator frequency $f=\omega/ \left( 2\pi \right)$ is shown in Fig.~\ref{plotesempio}. Amplitude and phase response of the RLC resonator are then fitted by modulus and phase of the complex transfer function of the circuit in Fig.~\ref{setup}(b), which can be expressed as: 
\begin{equation}\label{Vo}
V_o = \frac{Z_r}{Z_r+Z_c+\frac{1}{\frac{1}{R_p}+i\omega C_p + \frac{1}{Z_l}} } V_i ,
\end{equation}
where $Z_r=1/\left( i \omega C_r \right)$, $Z_c=1/\left( i \omega C \right)$, $Z_l=R+i \omega L = R_0 + \mathcal{R}+i\omega \left( L_0 + \mathcal{L} \right) $. 

{The resonant frequency of the circuit determined by Eq. (6) is approximately ${f_0=1/2 \pi\sqrt{L C_{eq}}}$, where the equivalent capacitance is ${C_{eq} \approx \left( C^{-1}+C_r^{-1}  \right)^{-1} + C_p} $. Thus we can tune $f_0$ by varying $C$ and $C_r$. $C_p$ represents the parasitic capacitance of the coil and $R_p$ the associated parallel dissipation. The value $C_p=310$ pF has been estimated in a separate measurement of the bare coil without capacitor together with the ohmic resistance $R_0=2.03$~k$\Omega$.}

{We performed measurements with 4 different sets of capacitors $C,C_r$, and thus $f_0$. This allows us to test our model over a wide range of parameters. For each set, we first estimate the inductance $L_0 \simeq 263$ H and the parasitic resistance $R_p\simeq 60$ M$\Omega$ by fitting the amplitude and phase of the output voltage with the rotor removed from the gap. Then, we insert the rotor and vary the rotor frequency $F=\Omega/2\pi$ over a range including values lower and higher than $f_0$. For each $F$ we fit again the amplitude and phase response, but now all parameters $C,C_r, C_p, R_p$ are kept as fixed parameters, and only $R$ and $L$, embodying the effect of the cylinder, are left as free parameters. 
The values of $R$ and $L$ extracted from amplitude and phase fits are typically consistent with each other, with relative discrepancies of the order of $10^{-4}$ for $L$ and $10^{-2}$ for $R$.}\\
\emph{Results} -- 
\begin{figure*}[t]
\centering
\includegraphics[width=2\columnwidth]{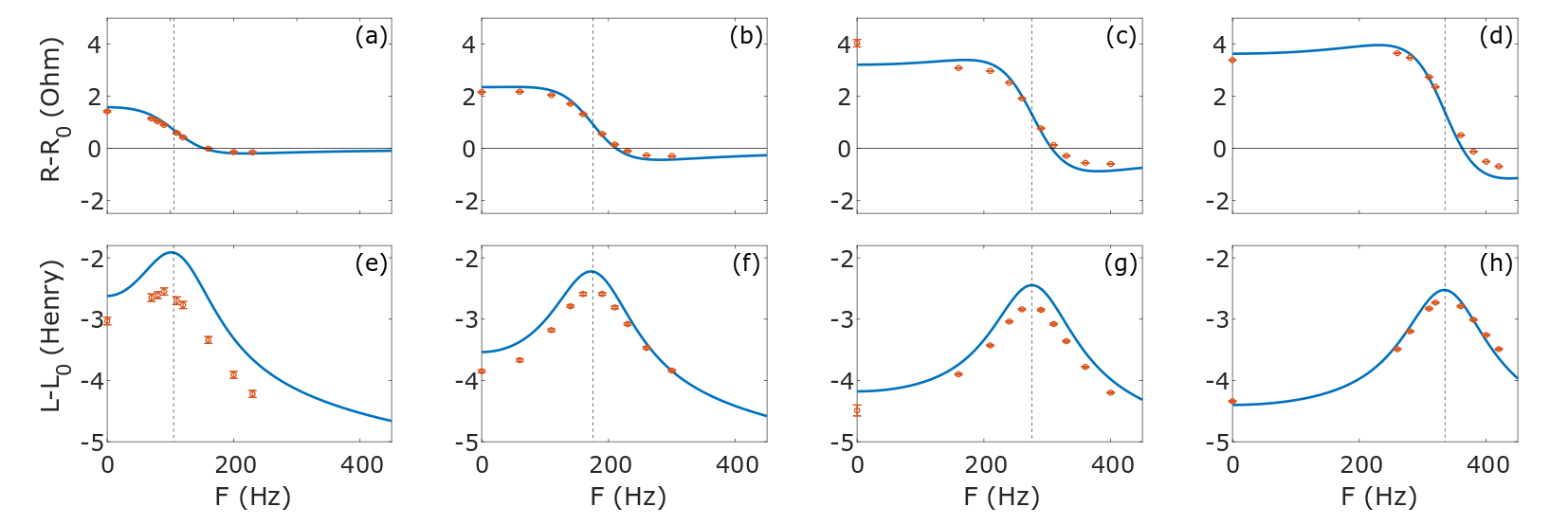}
\caption{\label{results} 
{Resistance and inductance induced by the rotor into the coil, $\mathcal{R}=R-R_0$ (a-d) and $\mathcal{L}=L-L_0$ (e-h), as function of the rotational frequency $F=\Omega/(2\pi)$ for different sets of capacitors, and hence different resonance frequencies of the LC resonator $f_0=\omega/2\pi$. Panels (a,e): $C=10$ nF, $C_r=47$ nF, $f_0 = 107$ Hz; Panels (b,f): $C=3.3$ nF, $C_r=22$ nF, $f_0 = 175$ Hz; Panels (c,g): $C=1.0$ nF, $C_r=22$ nF, $f_0 = 277$ Hz; Panels (d,h): $C=1.0$ nF, $C_r=1.0$ nF, $f_0 = 330$ Hz. The values obtained by the sphere model are shown as a reference (continuous lines). The vertical lines mark $f_0$. We notice that the negative values of $\mathcal{R}$ are negative by $6$ std. dev. at $f_0=107$~Hz and by $75$ std. dev. at $f_0=330$~Hz.}
}
\end{figure*}
Our best experimental estimations of the resistance and inductance induced by the rotor into the coil, $\mathcal{R}=R-R_0$ and $\mathcal{L}=L-L_0$, are shown in Fig.~\ref{results} as functions of the rotational frequency $F$ for the 4 different resonance frequencies $f_0$ of the LC oscillator. The continuous lines in the plots represent the theoretical curves for a rotating sphere. No free parameters were used to fit the model to the measurements, showing a substantial agreement with the experimental data. We note that the sphere-based model appears to capture more accurately border effects compared to the infinite cylinder model (see \cite{supply}). {A residual discrepancy is however expected. The larger discrepancy in the inductance data could be explained by parasitic coupling to the brush-less motor, which would be stronger at lower frequencies due to an increased penetration depth.} \\ 
\emph{Discussion} -- 
{According to Zel'dovich's original paper, amplification occurs due to two principal ingredients: 1) the Doppler shifted frequency has to become negative; 2) the imaginary (absorptive) part of the rotor material susceptibility changes sign due to the negative Doppler frequency, transforming losses into gain. }
Figure~\ref{results} shows that the experimental results follow the Zel'dovich trend for the resistance, and hence absorption, as a function of the rotor frequency \cite{BraidottiPRL2020}. When the rotor frequency $\Omega/2\pi$ exceeds the LC resonant frequency $f_0$, the co-rotating Doppler shifted frequency $\omega-\Omega$ and the corresponding resistance term become negative (note that $\sigma=1$ in our case), marking the inflection point in the resistance $\mathcal{R}$ plot. At slightly higher frequency the negative co-rotating term exceeds the positive counterrotating one, leading to a negative total reflected resistance $\mathcal{R}$. This is a signature that the absorption coefficient has flipped sign and hence that there is an electromagnetic gain induced by the mechanical rotation.\\
The effect is particularly evident at high resonance frequencies ($f_0=277$ {and} $335$ Hz) in agreement with the model. i.e. the imbalance between the negative resistance (gain) induced by the co-rotating component of the field and the positive resistance (absorption) induced by the counter-rotating component, increases with the magnetic field frequency $f$. 
We conclude that our experiment, based on the simplest possible interaction of a solid metallic cylinder with an oscillating magnetic field, is substantially reproducing the electromagnetic amplification mechanism predicted by Zel'dovich. {Furthermore, it is making a further step showing that this effect can be generalized to spin angular momentum.} 
The practical impossibility to test Zel'dovich predictions, pointed out already in the original paper \cite{Zeldovich1971}, is overcome here thanks to the heavy spatial confinement of the electromagnetic field in a LC resonator, compared to free space \cite{BraidottiPRL2020}, {and by making use of spin angular momentum}.\\
At the same time, {these results show an unexpected connection between the believed-inaccessible Zel'dovich effect and induction generators \cite{Keljik2009}, which extract electric power from rotational motion. 
We identify the Doppler shifted frequency with the `slip frequency' in induction motor terminology, and ingredient (1) is satisfied in the generator regime, when the rotor is driven faster than the rotating magnetic field induced by the stator excitation current. 
Our setup directly implements condition (2) with a homogeneous metallic rotor. However, typical induction generators are {made with} a squirrel cage rotor, i.e.\ circuits formed by multiple conducting bars around an iron core. Rather than the response of the rotor being determined by a solid material susceptibility, it is engineered through the resistances and inductances of the squirrel cage circuits. The analogue susceptibility of a squirrel cage enhances the amplification, for a more efficient generator than Zel'dovich's original proposal.}
\emph{Conclusions} --
By operating in the sub-wavelength regime {and using spin angular momentum, we have experimentally measured negative dissipation induced by a rotating metallic cylinder, indicating the amplification of EM waves originally predicted by Zel'dovich.} {This 60 year old prediction that was believed to be untestable with current technology is found to be observable thanks to the unforeseen link between Zel'dovich amplification and induction motors. These findings open} the way to the merging of ideas from two previously disconnected fields. 
{In particular,} a suggestive prospect is the realization of Zel'dovich electromagnetic amplification from a rotating body in the quantum regime \cite{Bekenstein1998, Maghrebi2012}, i.e. the generation of photons out of the quantum vacuum stimulated by a mechanical rotation \cite{Pendry_1997,Pendry2012,davies2005}. {In the past this seemed an impossible task, due to the extremely low efficiency of the effect and the need for having an EM resonator at frequency $\omega$ cooled to the ground state and combined with a rotor spinning at $\Omega>\omega$. }
{In contrast, induction motor schemes can be optimized through high efficiency magneto-mechanical coupling. It remains true that realistic values of $\omega, \Omega \sim 10^3$ Hz imply that the temperature required to bring the resonator in the ground state would be prohibitively low $T\sim 10^{-9} $ K. However, ground-state cooling of a low frequency LC resonator can be rather achieved using  techniques borrowed from optomechanics, such as feedback-cooling \cite{Vinante2010}. Such experiments would allow to observe of Zel'dovich amplification in the quantum regime. }


\begin{acknowledgments}
\emph{Acknowledgments} --
We thank Damon Grimsey for expert technical support with the setup. The authors acknowledge financial support from EPSRC (UK Grant No. EP/P006078/2) and the European Union's Horizon 2020 research and innovation programme, grant agreement No. 820392. 
A. V. and H.U. acknowledge financial support from the QuantERA grant LEMAQUME, funded by the QuantERA II ERA-NET Cofund in Quantum Technologies implemented within the EU Horizon 2020 Programme,  from the UK funding agency EPSRC (grants EP/W007444/1, EP/V035975/1 and EP/V000624/1), the Leverhulme Trust (RPG-2022-57), the EU Horizon 2020 FET-Open project TeQ (766900) and the EU Horizon Europe EIC Pathfinder project QuCoM (GA no.10032223). Data is available at: {DOI:10.5525/gla.researchdata.1457}
\end{acknowledgments}

%

\newpage
\onecolumngrid
\centering{\large \bfseries Amplification of electromagnetic waves by a rotating body: Supplementary material\par}\vspace{2ex}
	{ M. C. Braidotti$^{1}$, A. Vinante$^{2}$, M. Cromb$^{3}$, A. Sandakumar$^{3}$, D. Faccio$^{1,*}$, H. Ulbricht$^{3,*}$\par}
{\centering  \small \emph{$^{1}$School of Physics and Astronomy, University of Glasgow, G12 8QQ, Glasgow, UK.\\ 
$^{2}$Istituto di Fotonica e Nanotecnologie - CNR and Fondazione Bruno Kessler, I-38123 Povo, Trento, Italy.  \\
$^{3}$Department of Physics and Astronomy, University of Southampton, SO17 1BJ, Southampton, UK.\\
 $^*$}\par}
\smallbreak

\par\vspace{1ex}

\renewcommand{\theequation}{S\arabic{equation}}
\renewcommand{\thefigure}{S\arabic{figure}}
 \setcounter{equation}{0}

\section{The sphere model}
We model the system composed by the LC toroid circuit and the spinning cylinder within the toroid gap as shown in Fig. 1 of the main manuscript. Amplification occurs when the resistance $\mathcal{R}$ induced by the cylinder in the circuit becomes negative.
To compute the effect of the cylinder on the circuit we approximate the cylinder as a sphere with same radius $R$. This is justified by the fact that the gap is quasi-cubical, so the effective length of the cylinder inside the gap is close to $R$. Then, we compute the magnetic dipole moment $\bm{m_0}$ induced by the magnetic field in the conductive sphere. This will allow us to calculate the magnetic flux back-reflected into the circuit, and hence the induced resistance. To do this we follow the derivation in \cite{BraidottiPRL2020}, the difference lays in the magnetic field $\mathbf{B_0}$ which is oscillating rather than rotating. 

Let's consider the spinning axis oriented along $z$ and the magnetic flux density $\bm{B_0}$ produced by the coil wrapped around the toroid to be uniform and oscillating at a frequency $\omega$.\\
In the lab reference frame we can write the field in the toroid gap as: 
\begin{equation}
\mathbf{B_0}= \beta \mathbf{b_0}I,
\end{equation}
where $I$ is the current in the coil wrapped around the toroid. The geometrical factor $\beta= (0.40\pm 0.03$) T/A has been measured experimentally. 
The vector $\mathbf{b_0} = (1,0,0)^T e^{i\omega t}$ represents the linear polarization of the field oscillating at frequency $\omega$. To our purposes it is convenient to write it as the sum of two rotating vectors, co-rotating and counter-rotating with respect to the cylinder rotation, such that: 
\begin{equation}
\mathbf{b_0} ={1\over2} \left[(1,i,0)^T +(1,-i,0)^T \right] e^{i\omega t}. 
\end{equation}
In order to see the amplification effect we are looking for, calculations are carried out in the reference frame co-rotating with the rotor.
In this frame, we write the vector $\mathbf{b_0}$ as 
\begin{equation}
\mathbf{b_0} ={1\over2} \left[(1,i,0)^T e^{i(\omega-\Omega) t} +(1,-i,0)^T e^{i(\omega+\Omega) t}\right] = \mathbf{b}_{co} + \mathbf{b}_{counter},
\label{co_cout}
\end{equation}
where $\omega$ and $\Omega$ are the frequency of the field $\bm{B_0}$ and the rotation frequency of the rotor in the lab frame, respectively. \\
Our purpose is to find the form of the vector potential $\bm{\mathcal{A}}$, which allows us to compute the magnetic dipole moment, coupled to the magnetic flux $\phi=\beta|\bm{m_0}|$ back-reflected into the toroid circuit. 

To this aim, we solve Maxwell's equations, which describe the interaction of the magnetic field $\bm{B}$ generated by the circuit and the conductive sphere and we find the equation for the vector potential \cite{BraidottiPRL2020}:
\begin{equation}
-\Delta\mathbf{\mathcal{A}}+\mu_0\mu_r\sigma\frac{\partial\mathbf{\mathcal{A}}}{\partial t}=0.
\label{eqA}
\end{equation}
In order to solve this last equation for $\bm{\mathcal{A}}$, we divide the space in two regions: one inside the cylinder with radius $R$ and one outside it. At the interface of the two areas, we consider the following three boundary conditions for a spherical body:
\begin{eqnarray}
B_{\perp}^{in} &=& B_{\perp}^{out} \label{bc1}\\
H_{\parallel}^{in} &=& H_{\parallel}^{out} \label{bc2}\\
\mathbf{\mathcal{A}} \underset{r\rightarrow\infty}{\longrightarrow} &\mathbf{\mathcal{A}_{\infty}}&=\frac{1}{2}\mathbf{B_0}\times\mathbf{r} \label{bc3}
\end{eqnarray}
where $\mathbf{B}=\mu_0\mu_r\mathbf{H}$ and $\bm{r}$ is the position vector in the co-rotating reference frame centred at the center of the sphere. The magnetic field outside the sphere $\mathbf{B^{out}}=\mathbf{B_0}+\mathbf{B^{refl}}$ is composed of two components: the incident magnetic field generated by the circuit $\mathbf{B_0}$ plus the magnetic field reflected from the cylinder $\mathbf{B^{refl}}$. Let us remind that $\mathbf{B_0}$ is composed by a co-rotating and a counter-rotating component (see eq. (\ref{co_cout})). The magnetic field inside the cylinder $\mathbf{B^{in}}$ is given by the component of the incident field transmitted in the scattering.\\
Considering the last boundary condition, the complex vector potential $\mathbf{\mathcal{A}}$ can be generally written as 
\begin{equation}
\mathbf{\mathcal{A}} = \frac{1}{2}\left[\left(F_{co}(r)\mathbf{B_{co}}+F_{counter}(r)\mathbf{B_{counter}}\right)\times\mathbf{r}\right] = \mathbf{\mathcal{A}_{co}}+\mathbf{\mathcal{A}_{counter}},
\label{A}
\end{equation}
being $\mathbf{B_0} = \mathbf{B_{co}}+\mathbf{B_{counter}}$. The function $F(r)$, such that $F(r) \to \infty =1$, can be computed following \cite{BraidottiPRL2020} as \begin{eqnarray}
&&F(r) = 1+D(c_{EC}R)\left(\frac{R}{r}\right)^3 \qquad\mbox{for}\quad r\leq R\\
&&F(r) = \left[1+D(c_{EC}R)\right]\frac{f(c_{EC}r)}{f(c_{EC}R)}  \qquad\mbox{for}\quad r>R.
\end{eqnarray}
where $c_{EC}=\sqrt{-i\mu_0\mu_r\sigma\omega_{\pm}}$, with $\omega_{\pm} = \omega\pm\Omega$ depending whether we are considering the co-rotating component or the counter-rotating one. 
The function $D$ is
\begin{eqnarray}
&&D(c_{EC}R) = \frac{(2\mu_r+1)g(c_{EC}R)-1}{(\mu_r-1)g(c_{EC}R)+1}    \\
&&g(c_{EC}R) = f(c_{EC}R)\frac{c_{EC}R}{\sin(c_{EC}R)}=\frac{1-(c_{EC}R)\cot(c_{EC}R)}{(c_{EC}R)^2}.
\end{eqnarray}
These last equations define the explicit form of the vector potential $\bm{\mathcal{A}}=\mathbf{\mathcal{A}_{co}}+\mathbf{\mathcal{A}_{counter}}$ in the whole space. \\
It is now possible to compute the form of the magnetic dipole moment $\mathbf{m_0}$ on the cylinder in the reference frame co-rotating with the cylinder, which will be composed by two components: one co-rotating with the cylinder and another one counter-rotating. 
From the relation between $\mathbf{m_0}$ and the reflected component of the vector potential $\mathbf{\mathcal{A}}$
\begin{equation}
\label{AvsM}
\mathbf{\mathcal{A}}=\frac{\mu_0}{4\pi}\frac{\mathbf{m_0}\times\mathbf{r}}{r^3}
\end{equation}
we find
\begin{eqnarray}\label{magn_dipole}
\mathbf{m_0}&=&\frac{\pi}{\mu_0} I R^3 \beta [F_{co}(R)-1]\begin{pmatrix}
1 \\
i \\
0
\end{pmatrix} e^{i(\omega-\Omega) t} + \\
&+& \frac{\pi}{\mu_0} I R^3 \beta [F_{co}(R)-1]\begin{pmatrix}
1 \\
-i \\
0
\end{pmatrix} e^{i(\omega+\Omega) t}
= \\
&=&\left(\chi_{co} \mathbf{B_{co}}+\chi_{counter}\mathbf{B_{counter}}\right) .
\end{eqnarray}
where $\chi=\chi'-i\chi''$ is the complex response function of the sphere in the presence of the field in the co-rotating reference frame. It is worth noting that in order to compute the amount of reflected flux into the coil, we subtracted a factor $1$ from $F(R)$  corresponding to the contribution of the incident field $\mathbf{B_0}$. \\ 
Looking at the amplification from the point of view of the LC circuit, the rotating magnetic moment will couple with a magnetic flux $\phi$ back-reflected into the circuit and with peak value $\phi = \beta|\mathbf{m_0}|$.
We can thus write a relation $\phi = (\alpha_{co}+\alpha_{counter}) I$, with $\alpha = \alpha'-i\alpha'' $: the flux response function is given by $\alpha=\beta^2\chi$. 
The voltage given by the flux is then: 
\begin{equation}
V = i\omega\phi = i\omega (\alpha_{co}+\alpha_{counter}) I
\end{equation}
and we can define the resistance $\mathcal{R}$ and the inductance $\mathcal{L}$ generated by the presence of the rotating cylinder as:
\begin{eqnarray}
\label{m_yZ}
\mathcal{R} &=& Re[V/I] = \omega (\alpha''_{co}+\alpha''_{counter}) \\
\mathcal{L} &=& Re[\phi/I] = (\alpha'_{co}+\alpha'_{counter}).
\end{eqnarray}
The factor $\alpha'$ gives the inductance $\mathcal{L}$ generated by the presence of the rotating cylinder in the gap, while $\omega\alpha''$ can be seen as a resistance $\mathcal{R}$ induced by the rotating cylinder into the circuit. Hence amplification corresponds to a negative resistance that leads to power emission into the electromagnetic (EM) mode as opposed to the expected (for a non-rotating or slowly rotating cylinder) power absorbed from the EM mode.
We observe that the resistance $\mathcal{R}$ is composed by a co-rotating and a counter-rotating components. We have amplification only if the co-rotating one is larger than the counter-rotating one. This is possible since the response function $\alpha$ is not symmetric with respect to $\omega$.

\section{The cylinder model}

Let us consider a non rotating cylinder with infinite length in a transverse oscillating magnetic field, along $x$. 
To compute the effect of the cylinder we write the oscillating magnetic field back-reflected from the cylinder as a vector $\mathbf{B} = (B_r,B_{\phi},B_z)$ in cylindrical coordinates. Considering the geometry of our system the $B_z$ component is null since the external magnetic field is transverse to the cylinder
and no induced $\hat{z}$ component is anticipated. This leaves us with a 2D problem \cite{Perry1978}. Solving Maxwell equations for this problem gives the components of the back-reflected field from the cylinder:
\begin{eqnarray}
\label{Bout}
B_r &=& B_0 \frac{R'^2}{r^2}\cos(\phi)e^{-i\omega t} \\
B_{\phi} &=& -B_0 \Bigl(-\frac{R'^2}{r^2}\Bigr)\sin(\phi)e^{-i\omega t},
\end{eqnarray}
where $R'^2$ is given by
\begin{equation}
    R'^2(\omega) = R^2\frac{(\mu_r+1) J_1(\sqrt{iR_m}) - \sqrt{iR_m} J_0(\sqrt{iR_m}) }{(\mu_r-1) J_1(\sqrt{iR_m}) - \sqrt{iR_m} J_0(\sqrt{iR_m})}
\end{equation}
where $R_m(\omega) = \left[R/\delta(\omega)\right]^2$ is the magnetic Reynolds number and $\delta(\omega) = 1/ \sqrt{\mu_0\mu_r\sigma\omega}$ is the penetration depth, depending on the frequency of oscillation of the magnetic field $\omega$ \cite{Perry1978}.
Let us consider the case of rotating cylinder. Equations (\ref{Bout}) needs to account the rotation of the cylinder at frequency $\Omega$.
In the reference frame co-rotating with the cylinder, the co-rotating and a counter-rotating components of the oscillating magnetic field rotate at the frequencies $\omega_{-}=\omega - \Omega$ and $\omega_{+}=\omega + \Omega$ respectively. To include the rotation we observe that the field in eq. (\ref{Bout})  corresponds to a linearly polarized field expressed in polar coordinates \cite{Matsuo2011}
\footnote{The linearly polarized vector in $(x,y)$ coordinates \begin{equation}
\begin{pmatrix} 1\\ 0 \end{pmatrix}\end{equation}
corresponds to the vector
\begin{equation}
\begin{pmatrix}
\cos(\phi) \\
-\sin(\phi)
\end{pmatrix}\end{equation}  
in polar coordinates\cite{Matsuo2011}.}.
In polar coordinates, right and left rotating vectors can be written as 
\begin{equation}
|\mbox{right}\rangle=\frac{1}{\sqrt{2}}
\begin{pmatrix}
\cos(\phi)-i \sin(\phi) \\
-\sin(\phi) - i \cos(\phi)
\end{pmatrix}
\qquad \mbox{and}\qquad
|\mbox{left}\rangle=\frac{1}{\sqrt{2}}
\begin{pmatrix}
\cos(\phi)+i \sin(\phi) \\
-\sin(\phi) + i \cos(\phi)
\end{pmatrix}.
\end{equation}
We can hence express the field in (\ref{Bout}) as a sum of co-rotating (right) and counter-rotating (left) components. Importantly, in this frame the frequency of the co-rotating field is different from the frequency of the counter-rotating component, being $\omega_{-}=\omega - \Omega$ and $\omega_{+}=\omega + \Omega$ respectively. 
The back-reflected field for our system is
\begin{eqnarray}
\label{BfromEQ}
B_r &=& B_0 \Bigl[\frac{R'^2(\omega_-)}{2r^2}\Bigl(\cos(\phi)-i \sin(\phi)\Bigr)e^{-i\omega_- t} + \frac{R'^2(\omega_+)}{2r^2}\Bigl(\cos(\phi)+i \sin(\phi)\Bigr)e^{-i\omega_+ t}\Bigr]  \\
B_{\phi} &=& - B_0 \Bigl[\frac{R'^2(\omega_-)}{2r^2}\Bigl(-\sin(\phi)-i \cos(\phi)\Bigr)e^{-i\omega_- t} + \frac{R'^2(\omega_+)}{2r^2}\Bigl(-\sin(\phi)+i \cos(\phi)\Bigr)e^{-i\omega_+ t}\Bigr].
\end{eqnarray}
Similarly to the sphere case we now compute the equivalent magnetic moment induced in the cylinder. To this end, we write the $\bm{B}$ field generated by a linear distribution of magnetic dipole moment with density per unit length $\bm{\lambda}$ along the cylinder axis, that is:
\begin{equation}
    \mathbf{B}(r,\phi) = \int_\infty^\infty \frac{\mu_0}{4\pi} \Bigl( \frac{3\mathbf{r}(\bm{\lambda}\cdot\mathbf{r})}{r^5} - \frac{\bm{\lambda}}{r^3} \Bigr)dz,
\end{equation}
where $\mathbf{r}$ is the position vector in cylindrical coordinates. 
The solution of this equation is: 
\begin{equation}\label{BfromM}
    \mathbf{B}(r,\phi) = \Bigl(
\frac{\mu_0}{2\pi} \frac{\lambda_r}{r^2},
-\frac{\mu_0}{2\pi} \frac{\lambda_{\phi}}{r^2}
\Bigr)^T.
\end{equation}
Comparing eqs. (\ref{BfromM}) and  (\ref{BfromEQ}) we find the expression for the magnetic moment density $\bm{\lambda} = (\lambda_r,\lambda_{\phi})^T$ being
\begin{eqnarray}
\label{momentum}
\lambda_r &=& \frac{2\pi r^2}{\mu_0} \frac{1}{2 r^2}B_0 \Bigl[R'^2(\omega_-)\Bigl(\cos(\phi)-i \sin(\phi)\Bigr)e^{-i\omega_- t} + R'^2(\omega_+)\Bigl(\cos(\phi)+i \sin(\phi)\Bigr)e^{-i\omega_+ t}\Bigr]=\\ 
\lambda_{\phi} &=& \frac{2\pi}{\mu_0} B_0 \frac{1}{2} \Bigl[R'^2(\omega_-)\Bigl(-\sin(\phi)-i \cos(\phi)\Bigr)e^{-i\omega_- t} + R'^2(\omega_+)\Bigl(-\sin(\phi)+i \cos(\phi)\Bigr)e^{-i\omega_+ t}\Bigr].
\end{eqnarray}
Now, for the actual finite case we take the solution for the density of magnetic moment of the infinite case and integrate over the finite length of the gap $L$, where the $\bm{B}$ field is non negligible. This gives the total effective magnetic moment induced in the finite cylinder, neglecting border effects:

Being $\mathbf{m} = \chi_{co}\mathbf{B_{co}} + \chi_{counter}\mathbf{B_{counter}}$, we find 
\begin{eqnarray}
\label{momentum}
\chi_{co} &=& \frac{2\pi}{\mu_0} R'^2(\omega_-) L  \\
\chi_{counter} &=& \frac{2\pi}{\mu_0}R'^2(\omega_+) L.
\end{eqnarray}
The resistance $\mathcal{R}$ and inductance $\mathcal{L}$ generated by the presence of the rotating cylinder can be written as
\begin{eqnarray}
\label{m_yZ}
\mathcal{R} &=& \omega \beta^2(\chi''_{co}+\chi''_{counter}) \\
\mathcal{L} &=&  \beta^2(\chi'_{co}+\chi'_{counter}).
\end{eqnarray}
\begin{figure}[t]
\centering
\includegraphics[width=1\columnwidth]{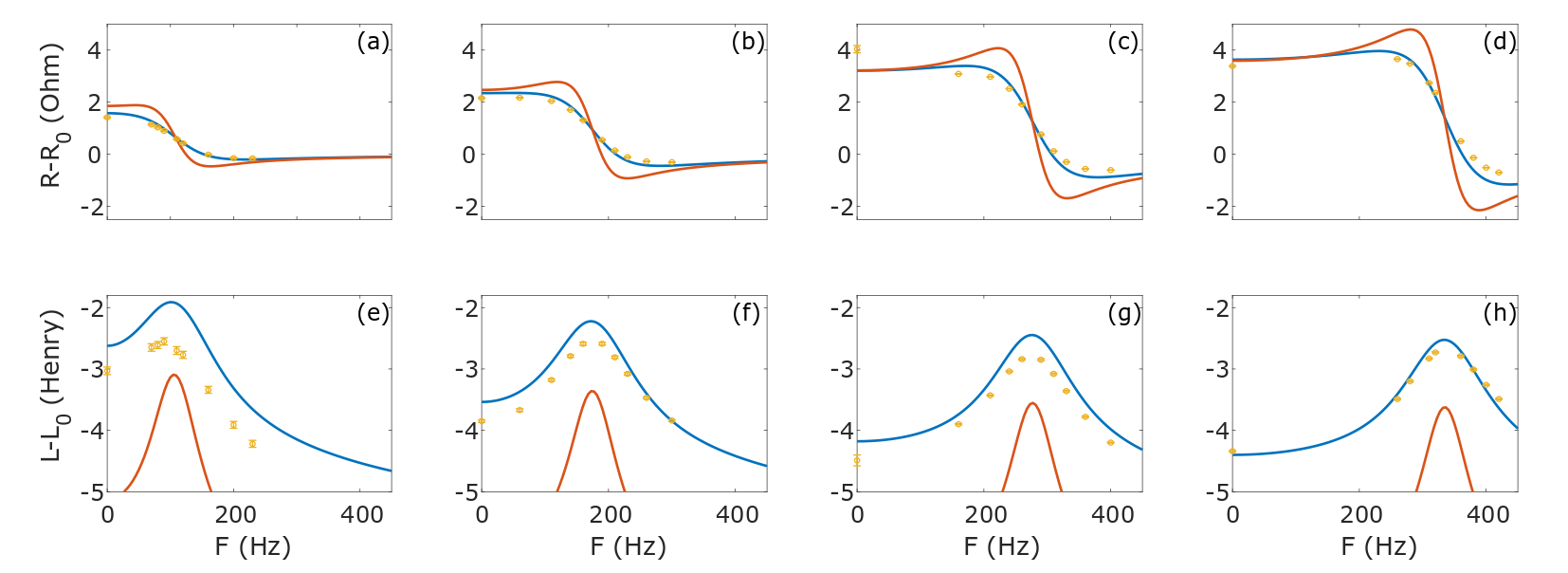}
\caption{\label{results} Comparison of the resistance  and inductance calculated through the sphere (blue) and cylinder (orange) models. $\mathcal{R}=R-R_0$ (a,b,c,d) and $\mathcal{L}=L-L_0$ (e,f,g,h) computed as function of the rotational frequency $F=\Omega/(2\pi)$ for different resonance frequencies $f=\omega/(2\pi)~=~\left[105 \textrm{(a,e)}, 175 \textrm{(b,f)}, 275 \textrm{(c,g)}, 335 \textrm{(d,h)}\right]$~Hz of the LC oscillator.    
}
\end{figure}

\section{Squirrel-cage induction generator vs solid rotor}
Induction machines are a type of rotating machine where a stator circuit - the non-rotating part, usually embedded in a hollow cylinder - has alternating currents oscillating at a single frequency $\omega_{ac}$ - the frequency of the power supply, which are used to generate a rotating field. The stator can have multiple magnetic poles $p$, hence the frequency of rotation of the stator magnetic field is $\omega_s = \omega_{ac}/p$.
\begin{figure}[h]
\includegraphics[width=0.5\columnwidth]{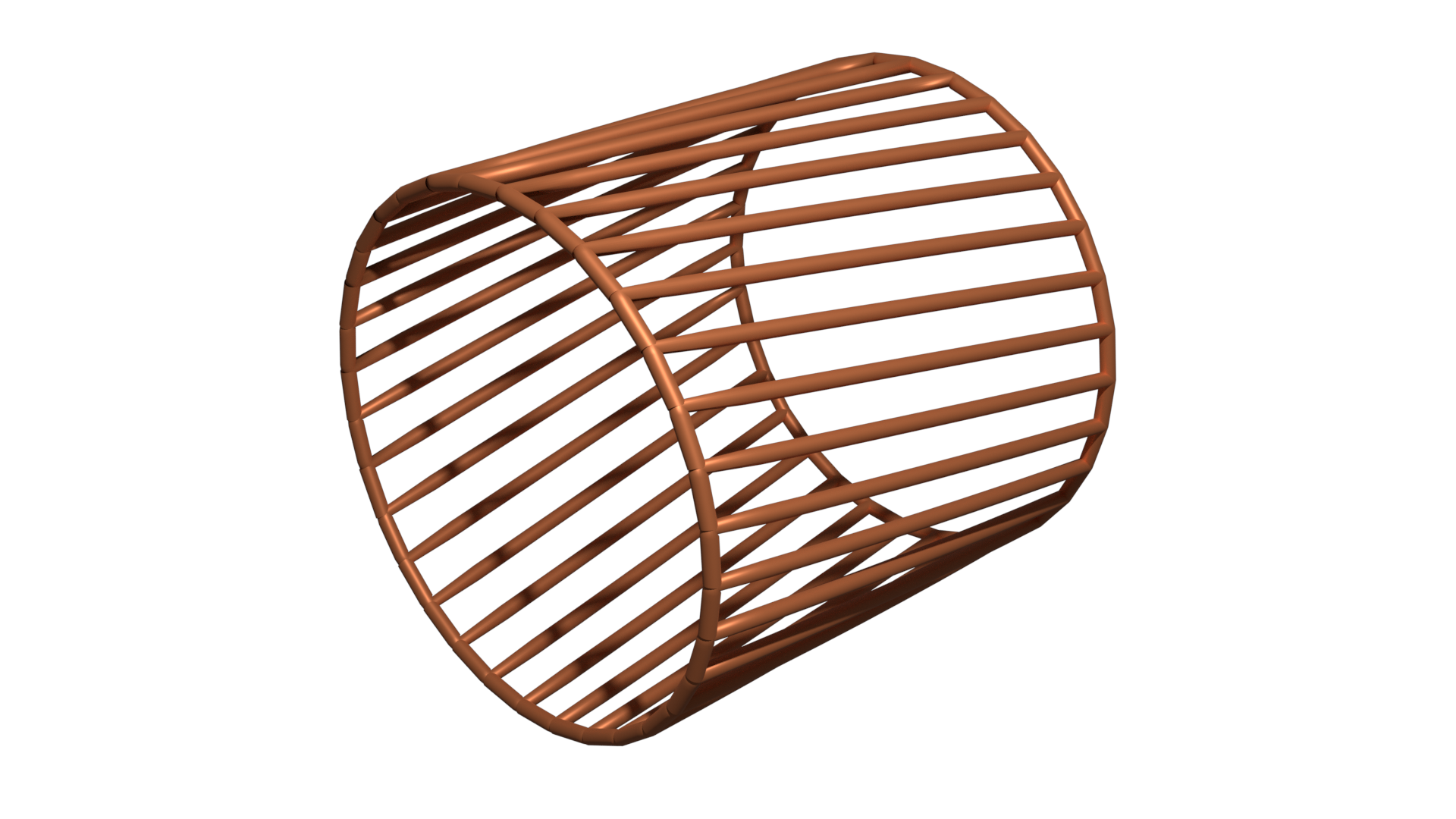}
\caption{\label{Squirrel_cage} Sketch of copper squirrel cage rotor.
}
\end{figure}
Secondary circuits are mounted on the rotor, placed inside the hollow stator cylinder, and rotate at frequency $\omega_m$ - the mechanical rotation. The mechanism is based on the principle of induction: the rotor circuits are short-circuited and their currents are induced by the stator rotating field. 

In most induction machines, a squirrel cage rotor is used. This is made of a solid cylinder of iron or steel carved by conductor bars, usually made of copper or aluminium, that are short-circuited by a conductive ring around the cylinder bases (see an sketch of the conductive bars in the typical squirrel cage circuit in Figure \ref{Squirrel_cage}). The thickness and the shape of these conductive bars determines the inductive and resistive properties of the rotor circuits. 
In our analysis we consider the induction system described in \cite{Woodson1968} and we write the voltage $V_s$ induced by the stator on the rotor as 
\begin{equation}
    V_s = \frac{\omega_s^2 M^2 (R_{rot}/s)}{(R_{rot}/s)^2 + \omega_s^2L_{rot}^2} I_s + i \frac{\omega_s^3M^2L_{rot}}{(R_{rot}/s)^2 + \omega_s^2L_{rot}^2} I_s,
\end{equation}
where $s = (\omega_s-\omega_m) / \omega_s $ is the so called slip, that can be related to the Zel'dovich condition considering the correspondence $p\rightarrow \ell$ ($\ell=1$ in our case), $\omega_m \rightarrow \Omega$ and $\omega_{ac}\rightarrow\omega$. $M$ is the mutual inductance, while $R_{rot},L_{rot}$ are the resistance and inductance of the rotor's circuits.\\
It is possible to find the effective susceptibility $\chi_{\mathrm{eff}}$ of the squirrel cage induction generator by recalling that $\alpha = \beta^2\chi_{\mathrm{eff}}$. The geometrical factor $\beta$ for the squirrel cage circuit is just the mutual inductance $M$. 
After some manipulation, we find:
\begin{equation}
    \chi_{\mathrm{eff}} = \frac{\alpha}{M^2} =  \frac{L_{rot}(\omega-\Omega)^2}{R_{rot}^2 + L_{rot}^2(\omega-\Omega)^2} - i \frac{R_{rot}(\omega-\Omega)}{R_{rot}^2 + L_{rot}^2(\omega-\Omega)^2} ,
\end{equation}
where the first term on the right hand side is the inductive component ($\chi_{\mathrm{eff}}'$) and the second is the resistive component ($\chi_{\mathrm{eff}}''$), which indeed flips sign if the slip frequency $(\omega-\Omega)<0$, i.e. the Zel'dovich condition is fulfilled.

\end{document}